# Safety of self-assembled neuromorphic hardware

**Can Rager***  
Clipco Ltd.

**Kyle Webster, PhD**  
Clipco Ltd.

## Summary

The scalability of modern computing hardware is limited by physical bottlenecks and high energy consumption. These limitations could be addressed by neuromorphic hardware (NMH) which is inspired by the human brain. NMH enables physically built-in capabilities of information processing at the hardware level. In other words, brain-like features bias hardware towards intelligence at scale. In Table 1 we compare computing devices by their ability to scale (scaling features) and adaptation of brain-inspired concepts (neuromorphic features). Neuropmorphic computing paradigms require a novel approach to safe, interpretable AI. In order to effectively engage the risk of misaligned AI, safety research may need to expand its scope to include NMH. This may be best achieved by supporting those currently engaged in NMH capability research to work on safety and related areas.

Table 1: **Comparison of computational hardware.** Comparison of the human brain [1]–[5] with neuromorphic hardware (atomic switching networks) [6]–[9] and a modern GPU (NVIDIA A100 PCIe 80GB) [10]. NMH shows higher energy efficiency than modern GPUs while also exhibiting brain-inspired features. Note that challenges in controlling nanowire networks currently limit their use cases [11]. Overcoming these challenges may increase our concern, as discussed in section 3 and 7.

|  | human brain | neuromorphic hardware | GPU |
|---|---|---|---|
| **scaling features** |  |  |  |
| node density | $2.3 \times 10^5$ /mm$^3$ | $10^7$ /mm$^2$ | $6.6 \times 10^7$ /mm$^2$ |
| signal rate (max.) | 338(10) Hz | $10^9$ Hz | $1.6 \times 10^9$ Hz |
| number of nodes | $8.6(8) \times 10^{10}$ | $1.6 \times 10^8$ | $5.42 \times 10^{10}$ |
| connections per node | $7 \times 10^3$ (per neuron) | 4 | 4 (per logic gate) |
| computational activity | $7.7 \times 10^8$ Hz / mm$^3$ | $10^{16}$ Hz / mm$^2$ | $1.1 \times 10^{17}$ Hz / mm$^2$ |
| energy consumption | $10^{-13}$ J/FLOP | $10^{-14}$ J/FLOP | $10^{-12}$ J/FLOP |
| **neuromorphic features** |  |  |  |
| self-assembly | yes | yes | no |
| plasticity | yes | under development | no |
| spike-based computing | yes | under development | no |
| criticality | yes | yes | no |
| memristive behavior | yes | yes | no |
| in-memory computing | yes | under development | no |

* Correspondence to can.rager@gmail.com

# 1. Brief introduction to nanowire networks

Biological brains have a spatially complex structure (fig. 1a). This structure inspired the design of computational hardware with highly interconnected nanowires (fig. 1b). The non-deterministic arrangement of wires is formed through *self-assembly*: Networks are exclusively shaped by microscopic interactions among individual nanowires [12].

The cross-point junctions between nanowires allow electrical signals to propagate through the network. Figure 1c shows a nanowire network placed on a chip with multiple readout ports. Feeding and measuring electrical signals across the network enables information processing. Furthermore, dynamic properties of nanowires enable short-term memory at the hardware level. Network plasticity (the ability to reorganize the network structurally and functionally through experience) is governed by two mechanisms [9]:

1. *Reweighting* by manipulating the conductivity of junctions (fig. 1d, 1e).
2. *Rewiring* by forcing a rupture or reconnection of single nanowires (fig. 1f, 1g).

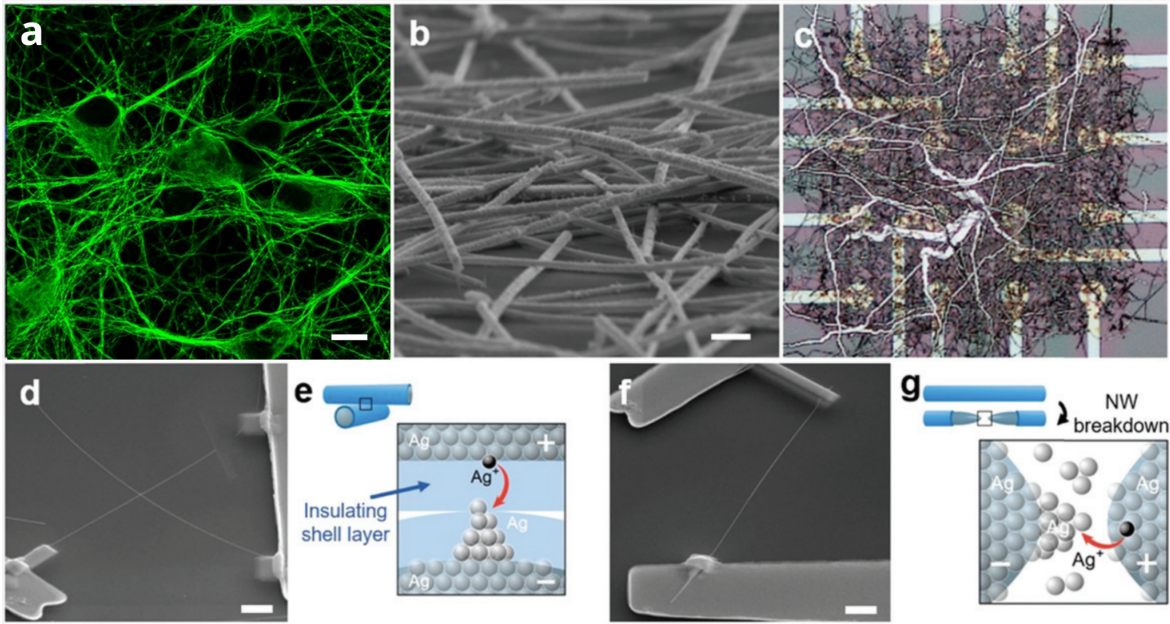

Figure 1: **Brain structure and nanowire networks. (a)** A biological neural network: Mouse hippocampus neurons stained for beta-III-tubulin after 6 days in culture (scale bar, 10 μm). Adapted from [13], [14]. **(b)** A biologically inspired memristive nanowire network (scale bar, 500 nm). Adapted from [9] under the terms of Creative Commons Attribution 4.0 License, copyright 2020, Wiley-VCH. **(c)** Atomic switch network of Ag wires placed on a chip with multiple readout ports. Adapted from [6], copyright 2013, IOP Publishing. **(d)** and **(e)** Single NW junction device where the memristive mechanism rely on the formation/rupture of a metallic conductive filament in between metallic cores of intersecting NWs under the action of an applied electric field and **(f)** and **(g)** single NW device where the switching mechanism, after the formation of a nanogap along the NW due to an electrical breakdown, is related to the electromigration of metal ions across this gap. Images and description adapted from [9], [11] under the terms of Creative Commons Attribution 4.0 License, copyright 2022, IOP Publishing.

To our knowledge training algorithms exclusively relying on *reweighting* and *rewiring* do not yet exist. In section 3 we argue control of plasticity is the main bottleneck of AI applications with nanowires. The systematic optimization of plasticity requires the following advances:

List 1: Current challenges in controlling plasticity of nanowires [6].

- Increasing the number of functional input/output ports in nanowire devices
- Designing interfaces (e.g. translating images to an electrical, time-dependent input signal)
- Selecting optimal materials capable of short-term and long-term memory
- Modeling of junction conductivity and switching behavior at scale
- Understanding the impact of scale-invariance and criticality on emerging intelligence

## 1.1    Potential use case: Efficient inference with nanowire devices

Nanowires have the potential to become the new standard for inference devices. We first present an existing training framework that may be applied suitable for nanowires. Then, we highlight why nanowires potentially outperform common hardware in fabrication cost, energy efficiency, and computation speed.

**Training framework.** Backpropagation has been successfully applied to well-understood physical systems already: Physics-aware training (PAT) enabled crystals and microphones to solve audio and image classification tasks [15]. Here, an external digital device feeds trainable parameters to the system as a separate input. The parameters are optimized with backpropagation. Once optimal parameters have been determined the external digital device is no longer required. The physical system is then able to perform inference on its own. As training execution is outsourced to a separate device, the physical system itself can be optimized for inference. Indeed, physical systems trained with PAT outperform inference with modern GPUs in energy efficiency and execution time [15].
Nanowire networks will become a suitable candidate for PAT once the issues in list 1 are better understood. Nanowires have been proven to outperform static systems in a comparable application (benchmark on image classification with reservoir computing [7]; section 7 provides more details on reservoir computing). Therefore, nanowires may become a primary candidate for PAT applications.

**Fabrication cost.** Common hardware like GPUs need to take the exact form of crossbar arrays to function properly. This specific design requires high manufacturing standards such as cleanroom facilities and high precision nano-lithography [16]. In contrast to that, self-assembled hardware such as nanowire networks do not require an explicitly predefined architecture [9]. Nanowire networks are created by casting a drop of nanowires in suspension on an insulating surface. This technique creates an unique arrangement of individual wires for each new device. Computational capability can be realized from many different morphologies as a starting point for network training. This indicates robustness to impure manufacturing. Therefore, self-assembly potentially enables cheaper production of hardware compared to common GPUs. This potential for low cost manufacturing also supports the case for use of nanowire technology in PAT applications.

**Energy efficiency and computation speed.** Spike-based computing within nanowire networks mimics the propagation of time-dependent signals through the human brain [17]. Here, a part of the input is encoded in signal timing. This allows an overall reduction of the input signal compared to deep neural networks [18]–[20]. The reduction of total signals yields a reduction in energy consumption. A higher number of connections per node further enables faster inference. Scaled spike-based networks have already been implemented in non-self-assembled NMH [21], [22]. Their application meets the state of the art whilst reducing energy consumption by two orders of magnitude [23]. There is a potentially important connection to the state of safety research here as we are not aware of interpretability research focused on spike-based computing. However, work on this problem could be undertaken by well placed capability researchers and may prove complementary to efforts in spike-based systems as undertaken by researchers focused on algorithms. We are not aware of work currently focused on either of these specific areas.

Secondly, in-memory computing is a key feature of NMH. The separation between memory and processing units is a standard in modern hardware design [24]. This architecture requires a data transfer between both units that limits the overall execution speed (also called the von-Neumann-bottleneck). Inspired by the human brain, in-memory computing overcomes this bottleneck. An ideal in-memory processor performs matrix vector multiplication with $\mathcal{O}(1)$ time complexity by taking advantage of analogue memory units [25] (compared to $\mathcal{O}(N^2)$ time complexity in common processors). This paradigm potentially increases the energy efficiency and computation speed by two orders of magnitude [26], [27]. Limited experimental progress has demonstrated the viability and direction of improvement for in-memory processors [28], [29].

## 2.  Mapping concerns onto the current state of technology

Advanced NMH presents a cluster of concerns. Throughout this section, we discuss interpretability and capability in the form of human disempowering AI.

**Interpretability.** Interpretability can significantly increase the probability that AI systems are beneficial to all stakeholders, particularly in complex applications. Uninterpretable models complicate auditing processes and may lead to unintended and human disempowering decisions.

***Further research on physical dynamics of NMH supports interpretability*** in our view. Understanding the large-scale dynamics of analog memory units is a necessary step towards model explainability. Nanowire networks are a particularly useful device to study: The self-assembled structure offers insight into scale-governed dynamics – also observed in mammal brains [30].

AI research on emergent behavior of large language models explicitly formulate scaling laws [31]–[33]. These laws could provide valuable insight for large-scale switching dynamics of NMH. We recommend closer alignment of NMH research to interpretability efforts (such as those led by Anthropic and Redwood Research).

**Misaligned artificial general intelligence.** In line with the scaling hypothesis [34], generalization and sophisticated behavior emerges from upscaling basic AI models with narrow capabilities. A growing field of researchers concerned by a potential emergence of artificial general intelligence (AGI) [35]. An AGI is able to define and execute instrumental sub-tasks to achieve objectives. By default, an AGI could seek power as an instrumental subtask to achieve almost any overall goal set by humans. Such a behavior might lead to disempowerment of humans and could be an existential risk to humanity [36].

Current nanowire networks exhibit dominant short-term memory [7]. This decreases the likelihood of long-term self-optimizing capabilities. However, this risk should be tracked by capabilities researchers as advances in hardware are both anticipated and realized. An effective response to emerging risks may require action prior to new technologies being realized, even in a relatively controlled research environment.

Whether neuromorphic features bias advanced AI towards **consciousness** remains an open question. The generalization capability of conscious systems potentially increases the risk of misaligned AGI. We recognise further exploration of these claims as a valid domain of inquiry but have ruled them out of scope of the current piece.

## 3. Potential developments that increase or decrease safety concern

Throughout this section we describe potential developments that would increase concerns discussed in section 2.

The potential development of **long-term memory** within nanowire networks would support feedback loops during the training process. This may increase the risk of misaligned AI. Further, the ability to **control nanowire plasticity** is central for realizing AI applications. As shown in table 1, the physical scalability of NMH is already comparable to modern GPUs. Once plasticity training is available, nanowire networks could become readily scalable hardware for AI applications. Advances in challenges (list 1) increase the likelihood of controllable nanowire networks.

We expect nanowire devices will undergo training with an external digital device (PAT) as discussed in section 1.2. This external training device may serve as a useful reference to further understand nanowire dynamics. The realization of **PAT with nanowires** and the performance and other properties of such systems would therefore provide an additional data source for AGI risk assessment.

Self-assembling networks could require lower manufacturing standards than classical computer chips [9]. The release of benchtop **manufacturing devices or 'kits'** for the production of self-assembled NMH by at first researchers, but later hobbyists would increase concern regarding advanced AI. Low-barriers to production would enable a broad audience to take part in the experimental development of self-assembled NMH. Accessible manufacturing potentially increases dual use risk as it accelerates the development of NMH and increases the risk of unilateral action. Additionally, this accessibility could worsen race dynamics towards advanced AI. This situation would complicate the monitoring of AI advances by governmental institutions or third parties. Analogous concerns are currently visible within the field of biosafety and 3D printed weapons.

**Criticality** may be an indicator for intelligence in nanowire networks. "The concept of criticality refers to a system poised at the point between ordered and disordered states (a 'critical point'), analogous to a phase transition." [30]. Multiple studies suggest critical dynamics may play a role in information processing and task performance in biological systems [37]–[39]. However, the relation between criticality and intelligence is not yet clear. Criticality applies to the resistance of nanowire networks. If the resistance is too high, an electrical signal will vanish soon after its release. Likewise, if the resistance is too low, a single signal can cause an avalanche of signals propagating through the whole network. The critical state of the network is positioned in the middle of these two extreme cases. The critical resistance of nanowire networks is optimal for information processing [7]. Advanced knowledge in critical systems would help to monitor capabilities of neuromorphic hardware.

## 4. Mapping AI alignment work onto neuromorphic networks

The most successful AI models among recent publications are massively scaled software architectures [40]–[42]. Naturally, the efforts of AI alignment are focused on large software architectures as well. However, a paradigm shift in computing hardware toward NMH could reshape the AI landscape in unexpected ways. This shift would be accelerated as modern computing hardware encounters physical bottlenecks while computation with physical systems possibly yields higher efficiency in execution time, energy, and data (section 3). Such a shift would require AI safety research to widen the software-focused scope to include hardware predictability.

In a recent review [11], members of industry and academia present the state-of-the-art in neuromorphic computing. They only dedicate a single paragraph within 97 pages to the emergence of misaligned AI as a concern. This suggests that AI safety is still neglected in neuromorphic computing research.

We argue investments in NMH are likely due to the prospects of higher energy efficiency, faster computation, and cheaper manufacturing. If industry funding grows, we expect profit maximizing incentives to favor prioritization of capability over safety [43]. Facing concerns in section 2, we recommend funding existing capability researchers to work on safety of NMH collaborating with the existing safety community where beneficial. This would ensure that work in domains such as hardware interpretability is supported by a deep technical knowledge base of the relevant systems, and that current safety researchers make only limited trade offs against urgent work in relatively mature AI software fields.

## 5. Counter Arguments

1. NMH is developing too slowly to matter. Other developments will determine the course of AGI prior to NMH becoming practical for real world tasks.

   *Short response:* We welcome work by technological forecasting experts to assess both the likelihood and impact of NMH capability becoming competitive with conventional approaches over relevant timelines for AGI. To our knowledge an assessment of this type has yet to be published.

2. We will potentially be able to simulate NMH with classical hardware. Thus, safety research of NMH is directly mappable to safety of classical hardware established research. NMH safety does not require explicit focus.

   *Short response:* The simulation of physical systems at scale is often not feasible. Additionally, the training framework PAT is not directly mappable to deep neural networks [15]. Even an ideal simulation of NMH would require novel interpretability methods.

3. We mention self-assembly mechanisms that enable live reorganization of a network as a key concern for AGI safety. However, adaptive plasticity is not supported by experimental results yet. Until that is the case, dedicating research effort to NMH is not worth the opportunity cost.

*Short response:* In line with our response to counter argument 1, we welcome an assessment of the cost-benefit for further work in this domain. We expect that the rate of progress on development of network re-organisation as described here is highly uncertain.

4. As of today, most neuromorphic networks are exclusively produced in specialized labs. During further research we could find major requirements in manufacturing that complicate the cheap and accessible production of self-assembling neuromorphic computing chips.

   *Short response:* This is likely to be the case where creation of networks strictly requires the use of expensive equipment or techniques such as electron microscopes or lithography. However, in the case of molecular biology there is a trend toward democratization of techniques previously only achievable by the most capable and well resourced labs. This process can start with a goal of facilitating replication of results by other scientists through 'productization' of techniques as realized through relatively easy to use, open source software and wet lab kits. Commercialisation of useful research technology potentially follows, finally making the jump into full accessibility when kits are openly sold to hobbyists. If nanowires prove not to require expensive manufacturing equipment at a useful or interesting stage of development, a similar process may occur.

# 6. Appendix

**Connections per node.** For the brain we use the average number of synaptic connections per neuron. For atomic switching networks we use the number of directly connected junctions per single junction. For GPUs we use the fan-out value per logic gate.

**Comparability of energy consumption among hardware.** The energy consumption of the human brain was calculated by $\frac{power\ consumption}{\#synapses\ *\ synaptic\ firing\ frequency\ *\ \#FLOP\ per\ synaptic\ operation}$ using 20 Watts of power consumption for an average brain [5].

For interconnected nanowire networks, only values for single input pulses of 10 ms duration (not repetitive spikes) are available.

The energy consumption per was calculated by $\frac{total\ energy\ per\ input\ pulse}{\#junctions}$. We expect a more comparable value once energy consumption data is available for nanowire networks with repetitive pulses. An isolated synaptic operation of an artificial synapse requires $10^{-15}$ Joule per operation [44]. This indicates potential for further increase in energy efficiency for nanowire networks.

A modern GPU consumes $10^{-12}$ Joule per floating point operation (FLOP).

The algorithms of NMH and modern computing hardware are not necessarily isomorphic (able to represent each other). Thus, the energy consumption of elementary operations of NMH and GPUs is only comparable w.r.t. a specific task. We compare the energy consumption w.r.t. an image classification task. López-Randulfe et al. [45] determine the number of elementary operations required to classify handwritten digits from the MNIST dataset. They benchmark both a Spiking Neural Network (SNN, a simulation of NMH) and a Convolutional Neural Network (CNN). We arrive at a conversion rate of 1 synaptic operation ≈ 10 FLOPs. Note, the conversion rate could vary for other benchmarks. We chose nanowires to represent NMH in Table 1, as they incorporate multiple neuromorphic features [11].

**Reservoir computing** utilizes the random network structure as a black box within a predictive algorithm [46]. Nanowire networks successfully perform image recognition with reservoir computing [7]. The combination of nanowires and shallow crossbar array chips successfully performs image recognition [7], speech recognition [47] and other tasks [6], [48]. We argue nanowires will likely develop as independent inference devices, initially trained by an external digital trainer (see physics-aware training below). Embedding the unknown network structure into AI poses a major challenge to model interpretability.

**Physics-aware training.** Intelligent systems are not bound to specifically designed hardware. Wright et al. [15] propose a general training algorithm for **arbitrary physical systems**, called physics-aware training (PAT). Applications include analog microphones and optical elements. While the algorithm uses a digital model for training, the main benefit of applying a physical system arises during inference: Execution time and energy consumption are decreased by multiple orders of magnitude compared to modern transistor-based hardware while achieving competitive accuracy.

**Representation learning** potentially upscales data and computational efficiency of NMH on an algorithmic level. A single data point is mapped to a specific representation within NMH. The number of possible representations grows exponentially with the number of hardware nodes. Research initiatives towards efficient information representations potentially decrease the demand of data (higher performance gain per data sample) and computation. Simultaneously, representation learning research advances hardware interpretability. Further, representation learning possibly has an impact on sensor design. It will define format requirements on sensor data output to enable efficient processing in NMH and tighten the feedback loop between sensors and computational hardware. We are not aware of representation learning efforts regarding neuromorphic hardware.